\documentclass[reprint,10pt,amsmath,amssymb,floatfix,aps,pra,superscriptaddress]{revtex4-1}

\usepackage[dvips]{graphicx}
\usepackage{dcolumn}
\usepackage{bm}
\usepackage{verbatim}
\usepackage[bookmarks=true, bookmarksopen=true, bookmarksnumbered=true, bookmarksopenlevel=1, colorlinks=true, citecolor=blue]{hyperref}
\makeatletter

\newcommand{\rom}[1]{\expandafter\@slowromancap\romannumeral #1@}
\makeatother
\begin{document}


\title{Dependence of intrinsic rotation reversals on collisionality in MAST}

\author{J. C. Hillesheim}
	\email{jon.hillesheim@ccfe.ac.uk}	
	\affiliation{CCFE, Culham Science Centre, Abingdon, Oxon OX14 3DB, United Kingdom}
\author{F.I. Parra}
	\affiliation{Rudolf Peierls Centre for Theoretical Physics, University of Oxford, Oxford, United Kingdom}
	\affiliation{CCFE, Culham Science Centre, Abingdon, Oxon OX14 3DB, United Kingdom}
\author{M. Barnes}
	\affiliation{Rudolf Peierls Centre for Theoretical Physics, University of Oxford, Oxford, United Kingdom}
	\affiliation{CCFE, Culham Science Centre, Abingdon, Oxon OX14 3DB, United Kingdom}
\author{N.A. Crocker}
	\affiliation{University of California, Los Angeles, Los Angeles, California 90095, USA}
\author{H. Meyer}
	\affiliation{CCFE, Culham Science Centre, Abingdon, Oxon OX14 3DB, United Kingdom}
\author{W.A. Peebles}
	\affiliation{University of California, Los Angeles, Los Angeles, California 90095, USA}
\author{R. Scannell}
	\affiliation{CCFE, Culham Science Centre, Abingdon, Oxon OX14 3DB, United Kingdom}
\author{A. Thornton}
	\affiliation{CCFE, Culham Science Centre, Abingdon, Oxon OX14 3DB, United Kingdom}
\author{the MAST Team}
	\affiliation{CCFE, Culham Science Centre, Abingdon, Oxon OX14 3DB, United Kingdom}

\date{\today}

\begin{abstract}
Tokamak plasmas rotate even without external injection of momentum.  A Doppler backscattering system installed at MAST has allowed this intrinsic rotation to be studied in Ohmic L-mode and H-mode plasmas, including the first observation of intrinsic rotation reversals in a spherical tokamak.  Experimental results are compared to a novel 1D model, which captures the collisionality dependence of the radial transport of toroidal angular momentum due to the effect of neoclassical flows on turbulent fluctuations. The model is able to accurately reproduce the change in sign of core toroidal rotation, using experimental density and temperature profiles from shots with rotation reversals as inputs and no free parameters fit to experimental data.  
\end{abstract}

\maketitle
\section{Introduction}
Tokamak plasmas rotate intrinsically, even when there is no external input of momentum, and this intrinsic rotation can spontaneously change direction with relatively small changes in plasma conditions~\cite{bortolon_observation_2006,rice_rotation_2011,angioni_intrinsic_2011}.  In present day magnetic confinement fusion experiments, large toroidal flows are driven by injecting high energy neutral particle beams into the plasma core.  This rotation can stabilize large scale instabilities like resistive wall modes~\cite{garofalo_sustained_2002}, and its shear can suppress small, gyroradius scale instabilities~\cite{burrell_effects_1997}, reducing the loss of heat and particles due to turbulent transport.  Neutral beams are not expected to drive the same magnitude of toroidal flow in future experiments, like ITER, or in envisioned reactors, due to their high densities, large sizes, and higher energy neutral beams, so intrinsic rotation represents a potentially attractive replacement.  Understanding the origin of this intrinsic rotation is necessary for predictions of rotation in future tokamaks.

Intrinsic rotation reversals have been reported  in conventional tokamaks with only $\sim$10\% changes to line-averaged density~\cite{bortolon_observation_2006,duval_bulk_2007,rice_rotation_2011,rice_observations_2011,angioni_intrinsic_2011}.  Rotation measurements were acquired in this work with a Doppler backscattering (DBS) system~\cite{hillesheim_mastdbs_sub} that was installed on the Mega Amp Spherical Tokamak (MAST)~\cite{lloyd_overview_2003} and used for the first observation of this phenomenon in a spherical tokamak, demonstrating that rotation reversals are a generic and robust property of momentum transport across a variety of tokamak configurations.  This striking behavior of a sign reversal provides a strong test to challenge proposed mechanisms for momentum transport.  As in other devices, reversals occur at a line-averaged density that scales linearly with plasma current.  A database of Ohmic MAST plasmas with good DBS data has been compiled spanning line-averaged density $1.0 <  \left< n_e \right>/ 10^{19}\  \mathrm{m}^{-3} < 4.0$ and plasma current 400 kA $< I_p <$ 900 kA at on-axis toroidal magnetic field $B_{\phi}=0.5$ T.

The database allows theoretical expectations for intrinsic rotation reversals and momentum transport to be tested.  Intrinsic rotation is the result of a turbulent redistribution of momentum within the plasma, which can be driven by a number of effects~\cite{gurcan_2007, pat_2008, camenen_transport_2009, parra_turbulent_2010, parra_sources_2011, waltz_gyrokinetic_2011, camenen_consequences_2011, barnes_intrinsic_2013-1, sung_toroidal_2013, lee_turbulent_2014, lee_effect_2014}. We focus on the effect that non-Maxwellian components of the ion distribution function have on the turbulence \cite{barnes_intrinsic_2013-1}, which on theoretical grounds is always expected to be important in up-down symmetric configurations~\cite{felix_arxiv_2014}. Non-linear turbulence simulations investigating this effect have shown that the momentum flux can change sign when a collision frequency threshold is crossed (with no change to the linear turbulence drive). Collisionality has previously been identified in experiments as an important parameter for understanding Ohmic rotation reversals~\cite{rice_ohmic_2012,reinke_density_2013,mcdermott_core_2014}. We develop a new heuristic model based on \cite{barnes_intrinsic_2013-1} for the purpose of assessing whether this is an important mechanism in MAST, and show that it reproduces many characteristics of experimental measurements of rotation reversals.  

\section{The Experiment}
MAST is a spherical tokamak with $R_0 \approx 0.95$ m and $a \approx 0.6$ m.  DBS is a refraction-localized scattering technique, which provides local measurements of density fluctuations and plasma flows.  Cross-diagnostic comparisons showed good agreement for velocity measurements in plasmas where heating neutral beams were used, and charge exchange recombination spectroscopy measurements were possible~\cite{hillesheim_mastdbs_sub}, except inside of internal transport barriers (which are absent in the Ohmic plasmas studied here).  The 16 channel DBS system at MAST enabled rotation profiles throughout rotation reversals to be measured.  Measurements far into the core, at square root normalized poloidal flux $\sqrt{\psi}\approx 0.4-0.5$, were often possible with high-k trajectories, typically obtained at $k_{\bot} \rho_i \sim 10$, so direct measurements of low-k ITG or TEM (ion temperature gradient or trapped electron mode) turbulence are not available.  

\section{Observations of Rotation reversals}
\begin{figure}[!htbp]
\includegraphics[width=8.5 cm]{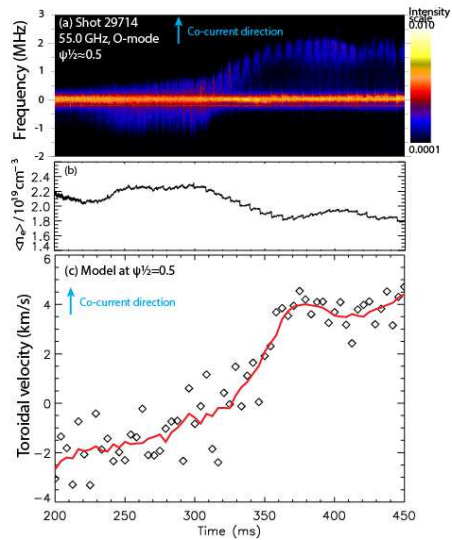}
\caption{\label{fig:29714_traces} (Color online) (a) DBS scattered electric field spectrogram measured at $\sqrt{\psi} \approx 0.5$, showing a large change to core rotation. (b) Line-averaged density.  (c) Model prediction for toroidal rotation at $\sqrt{\psi} = 0.5$.  Each diamond is a separate calculation corresponding to a different Thomson scattering measurement and the line is a boxcar average over $\sim 20$ ms.}
\end{figure}

Figure~\ref{fig:29714_traces}(a) shows the measured DBS electric field spectrum during a time when a large change to core intrinsic rotation is observed in an $I_p=$400 kA plasma.  The narrow, large amplitude signal near zero frequency that varies little with time is due to unlocalized high-$k_r$ backscattering and is not of interest here.  The broader, lower amplitude signal which starts close to zero or at slightly negative frequency, then increases is the localized DBS $k_{\bot}$ signal from the core plasma.  The Doppler shift frequency is due to the combination of the $E \times B$ velocity and turbulence phase velocity, $\omega_{DBS}=k_{\bot} v_{turb}=k_{\bot}\left(v_{E \times B} + v_{phase} \right)$, where the latter is typically much smaller than the former; the lab frame velocity of the turbulence is $v_{turb}$.  The scattering wavenumber of the density fluctuations, $k_{\bot}$, is determined via ray tracing.  When the toroidal flow far exceeds the diamagnetic velocity the radial electric field is dominated by toroidal rotation (this is almost always the case in the core of NBI-heated plasmas, but is not always true in Ohmic plasmas, particularly during reversals; see discussion in Sec.~\ref{sec:compare} for how this was accounted for in our comparisons).  We then estimate the toroidal rotation using $v_{\phi} \approx v_{turb} B/B_{\theta}$, where $B_{\theta}$ is the poloidal magnetic field and $B$ is the total field.  This yields $v_{\phi} \approx 0$ km/s at 300 ms in Fig.~\ref{fig:29714_traces}(a), increasing to $v_{\phi} \approx 30$ km/s at 400 ms (see later for full profile and uncertainties), which corresponds to a toroidal Mach number of $M_{\phi}=R_0 \Omega_{\phi}/v_{ti} \approx 0.1$ ($\Omega_{\phi}$ is the toroidal rotation frequency, $v_{ti}=\sqrt{2 T_i/m_i}$ is the ion thermal speed).  Figure~\ref{fig:29714_traces}(b) is the line-averaged density measured with an interferometer, which was feedback-controlled in this shot, showing the rotation changes on the same time scale as the density.   

\section{1D Intrinsic Rotation Model}
{In the absence of momentum injection, the total steady-state momentum flux is zero. To start, we assume that the intrinsic momentum flux, $\Pi_{int}$, is balanced by turbulent diffusion,
\begin{equation} \label{eq:totalPi}
m_i n_i \chi_{\phi} R^2_\psi \frac{\partial \Omega_{\phi}}{\partial r} = \Pi_{int},
\end{equation}
where $\chi_{\phi}$ is the momentum diffusivity and $R_\psi = \sqrt{\left< R^2 \right>}$ is the flux surface averaged major radius. We have excluded the turbulent pinch of momentum \cite{peeters_toroidal_2007} in Eqn.~\ref{eq:totalPi} for simplicity (it was a small effect in other Ohmic intrinsic rotation studies~\cite{angioni_intrinsic_2011} and cannot generate a reversal but only amplify flow generated by other mechanisms).  The turbulent momentum and energy diffusivities, $\chi_\phi$ and $\chi_i$, are related by the turbulent Prandtl number $P_r= \chi_{\phi}/\chi_{i}$, with typical value $P_r \approx 0.7$ \cite{casson_anomalous_2009, barnes_turbulent_2011} used in the model.}

{We want to identify contributions to $\Pi_{int}$ that reverse sign when the density changes. We focus on the turbulent momentum flux driven by the non-Maxwellian piece of the distribution function -- the neoclassical piece -- which is the result of the interaction between finite orbit widths, density and temperature gradients, and collisions~\cite{hinton_theory_1976,helander_collisional_2005}.  The turbulent momentum flux driven by it \cite{parra_turbulent_2010, parra_sources_2011, lee_turbulent_2014, lee_effect_2014} has been observed to reverse when the normalized collisionality $\nu_*= q R_\psi \nu_{ii}/ (v_{ti} \epsilon^{3/2})$ crosses a threshold $\nu_c \sim 1$~\cite{barnes_intrinsic_2013-1} because the neoclassical distribution function changes appreciably at the transition between the low collisionality (banana) regime and the intermediate collisionality (plateau) regime~\cite{hinton_theory_1976,helander_collisional_2005}. Here $\epsilon=r/R_\psi$ is the inverse aspect ratio of the flux surface, $q$ is the safety factor, and the ion-ion collision frequency is  $\nu_{ii} = 4 \pi n_i e^4 \ln \Lambda/ (\sqrt{m_i} (2 T_i)^{3/2})$,  where $\ln \Lambda$ is the Coulomb logarithm. To compare the predictions of \cite{barnes_intrinsic_2013-1} to experiments, we develop a simple heuristic model.}

{Fig. 1 of ~\cite{barnes_intrinsic_2013-1} gives the dimensionless measure of momentum redistribution $(v_{ti}/ R_\psi) (\Pi_{int}/Q_i)$ as a function of $\nu_*$ for ``cyclone base case''~\cite{dimits_comparisons_2000} parameters, where $Q_i$ is the turbulent energy flux.  To generalize the results of \cite{barnes_intrinsic_2013-1} we use scaling arguments to deduce the dependence of $\Pi_{int}$. The characteristic size of the neoclassical part of the distribution function is determined by the width of the particle orbits, $(B/B_\theta) \rho_i$, where $\rho_i$ is the ion gyroradius. We then expect corrections of order $(B/B_\theta) (\rho_i/L_{T_i})$, and in particular, intrinsic rotation levels of order $\Omega_\phi \sim (B/B_\theta) (\rho_i/L_{T_i}) (v_{ti}/R_\psi)$ with a characteristic radial gradient $\partial/\partial r \sim 1/L_{T_i}$, where $1/L_{T_i} = - \partial  \ln T_i/ \partial r$ is the inverse ion temperature gradient length scale.  With the above estimates we obtain the following model for the local momentum transport:
\begin{equation} \label{eqn:flux}
\frac{v_{ti}}{R_\psi} \frac{\Pi_{int}}{Q_i}  =  \frac{B}{B_{\theta}} \frac{\rho_i}{L_{T_i}} \widetilde{\Pi}(\nu_*),
\end{equation}
where $\widetilde{\Pi}$ is an order unity function that depends on $\nu_*$. We choose a form for $\widetilde{\Pi} (\nu_*)$,
\begin{equation}
\widetilde{\Pi} (\nu_*) = \frac{\widetilde{\Pi}_0 \left(\nu_*/\nu_c -1  \right)}{1 + (\nu_*/\nu_c) (\widetilde{\Pi}_0/\widetilde{\Pi}_{\infty})},
\end{equation}
which captures the main features of $\Pi_{int}$: there is a reversal at $\nu_* = \nu_c$, and the momentum flux does not depend on $\nu_*$ at sufficiently large and small values of $\nu_*$. The intrinsic rotation profile is found by solving for $d \Omega_\phi/dr$ in Eqn.~\ref{eq:totalPi}, using the relation $Q_i=n_i T_i \chi_i /L_{T_i}$, and integrating, 
\begin{equation} \label{eqn:model_vphi}
\Omega_{\phi}(\rho)= - \int^1_{\rho}  \frac{v_{ti} \rho_{*,\theta}}{2 P_r L_{T_i}^2}  \widetilde{\Pi}(\nu_*) d \rho + \Omega_{\phi}(1), 
\end{equation}
where $\rho$ is a flux surface label and $\rho_{*,\theta}=(\rho_i / R_{\psi})(B / B_{\theta})$.  To limit additional assumptions, we take $\Omega_{\phi}(1)=0$ rad/s, and estimate $\widetilde{\Pi}_0 = 0.3$, $\widetilde{\Pi}_{\infty}=1$, and $\nu_c=1.7$ from Fig.1 of \onlinecite{barnes_intrinsic_2013-1}, where $\widetilde{\Pi}_{\infty}$ is an extrapolation.  This results in a model that can use experimental temperature and density profiles and magnetic equilibrium geometry as inputs and predicts an intrinsic rotation profile.  Note that the model gives the local flux in terms of the local collisionality, so later organization of the data by the global parameters $I_p$ and $ \left< n_e \right>$ is only approximate.}

\section{Comparison between Model and Experiment}
\begin{figure}[!htbp]
\includegraphics[width=7.5 cm]{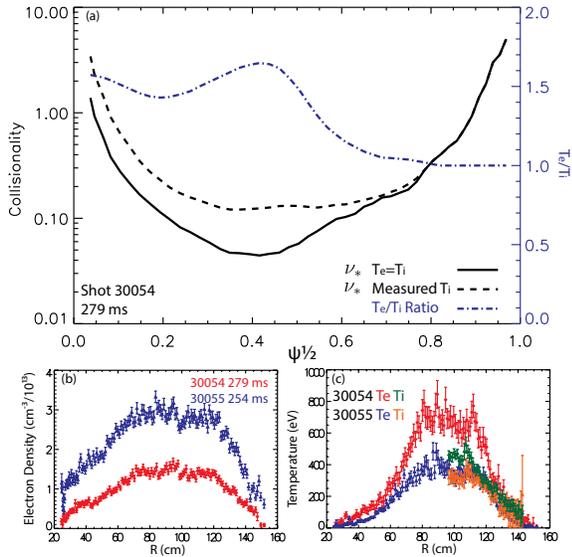}
\caption{\label{fig:teti} (Color online) (a) Effect of assuming $T_e=T_i$ on collisionality in shot 30054 at $I_p$=400 kA and $ \left< n_e \right>=1.0 \times 10^{13} \ \mathrm{cm}^{-3}$. (b) Electron density from 30054 at 279 ms (red diamonds) and 30055 at 254 ms (blue triangles).  (c) From the same time,  electron temperature in 30054 (red diamonds) and 30055 (blue triangles), and ion temperature in 30054 (green diamonds) and 30055 (orange triangles).}
\end{figure}

For these Ohmic plasmas, where no core ion measurements are usually available, we assume $T_e = T_i$ and $n_e=n_i$ (typically for the core plasma in MAST $Z_{eff} \lesssim 1.2$), and use Thomson scattering~\cite{scannell_130_2010} for $T_e$ and $n_e$.  Most shots in the database were purely Ohmic, but two did have short duration (10 ms) NBI blips for charge exchange recombination spectroscopy~\cite{conway_high-throughput_2006}.  Figure~\ref{fig:teti}(a) shows the effect of assuming $T_e=T_i$ on collisionality in one of these plasmas, using smoothed profiles; this case is actually the lowest density in the entire database ($I_p$=400 kA and $ \left< n_e \right>=1.0 \times 10^{13} \ \mathrm{cm}^{-3}$), so differential temperatures would be expected to be larger than most conditions.  We see that $T_e \approx T_i$ for $\nu_*\gtrsim 0.1-0.2$ (examination of the first 5 ms after NBI is applied for a variety of $I_p$ and $ \left< n_e \right>$ yields the same conclusion).  We have examined the results using either $T_e=T_i$ or using the measured $T_i$, and while the assumption of $T_e = T_i$ can impact the predicted magnitude of rotation (mostly through  differences in $L_{T_i}$ in this case), it does not affect the predicted sign since the differences are only large when the collisionality is small. The radii where $T_e=T_i$ is a reasonable approximation also largely coincides with the DBS measurement radii.  We therefore conclude this is a justifiable assumption for testing predictions of the sign of toroidal rotation.  Figure~\ref{fig:teti}(b-c) shows the measured density and temperature profiles for the two shots in the database with diagnostic beam blips, during the current flat top.  The $T_i$ measurements are averaged over 5 ms, during the NBI blips.  For the second shot, 30055 ($I_p$=400 kA and $ \left< n_e \right>=2.2 \times 10^{13} \ \mathrm{cm}^{-3}$), $T_e=T_i$ within uncertainties for the whole profile.  All profiles were measured close to the midplane.

Given the model's approximations, we restrict our comparisons to its robust features and do not expect detailed agreement on a case-by-case basis.  Figure~\ref{fig:29714_traces}(c) shows the model prediction for intrinsic rotation at $\sqrt{\psi}=0.5$.  The model predicts a change in core toroidal rotation towards the co-current direction (ion diamagnetic direction) at the same time as the experimental measurement, but the magnitude of the rotation is under-predicted.  Since the model parameters correspond to the cyclone base case, the pinch has been neglected, and the assumption of $T_e=T_i$ may affect core $L_{T_i}$, and we have assumed $\Omega_{\phi}(1)=0$ rad/s, the disagreement in magnitude is perhaps not surprising.  However, the crucial aspect of the model is that the momentum transport can change sign only near $\nu_* \approx 1$ (physically due to the transition from the plateau to the banana regime), which is a robust constraint that is only weakly sensitive to the exact number for $\nu_c$ since $\nu_*$ changes orders of magnitude across the plasma minor radius.  Independent of the under-prediction for magnitude, reversals of sign can therefore still be meaningfully compared.

\begin{figure}[!htbp]
\includegraphics[width=7.5 cm]{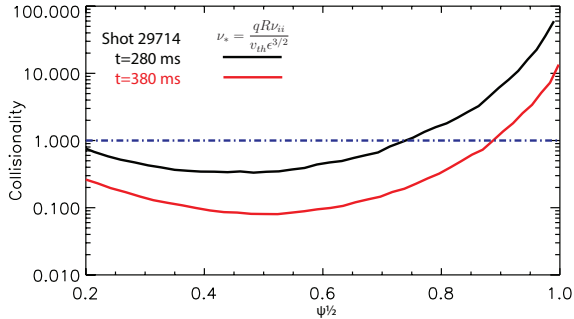}
\caption{\label{fig:29714_coll} (Color online) Collisionality profile before and after the rotation transition in 29714.  Reference line at $\nu_*=1$.}
\end{figure}

The collisionality change associated with the rotation transition is plotted in Fig.~\ref{fig:29714_coll}.  Note that $\nu_*$ is very large in the edge of Ohmic L-mode (no edge transport barrier) MAST plasmas, due to low temperature, which is about 10 eV at the separatrix for the 400 kA shots; typically in an H-mode plasma (plasmas with an edge transport barrier), $\nu_* \lesssim 1$ everywhere.  The factor that controls the sign of the rotation in the model in Eqn.~\ref{eqn:model_vphi} is the radial position where $\nu_* = \nu_c$.  The lower density at t=380 ms moves this point towards the edge, making the rotation gradient contribute to co-current rotation over most of the profile, yielding the change displayed in Fig.~\ref{fig:29714_traces}(c).  

\begin{figure}[!htbp]
\includegraphics[width=7.25 cm]{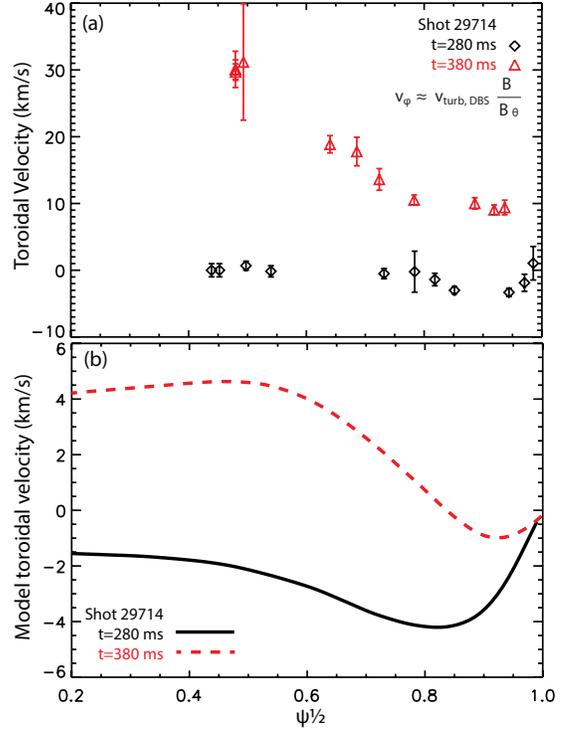}
\caption{\label{fig:29714_profile_mvx} (Color online) Comparison of (a) approximated experimental toroidal rotation profile and (b) model prediction for intrinsic rotation profile. }
\end{figure}

Figure~\ref{fig:29714_profile_mvx} compares the toroidal rotation profile approximated from DBS measurements (error bars do not include systematic uncertainties, discussed below in Sec.~\ref{sec:compare}) to profiles predicted by Eqn.~\ref{eqn:model_vphi}.  The model correctly predicts a large change towards co-current rotation as the density is decreased.  The experimental profiles are notable in that the rotation at the edge also changes significantly.  Typically in other experiments, even when the core toroidal rotation changes direction, the edge changes little.  Also notable is the small well near the edge at t=280 ms in both the experimental and model profiles.  The model always produces a feature like this unless $\nu_*<\nu_c$ for all radii.

We have so far focused on model predictions using parameters from available simulations, but limited investigations on the effects of changing model parameters have been conducted.  After accounting for the boundary condition by adjusting $\Omega_{\phi}(1)$, an increase of $\widetilde{\Pi}_0$ by about a factor of 3 is required to replicate the change in the core toroidal velocity at $\sqrt{\psi} \approx 0.5$ in Fig.~\ref{fig:29714_profile_mvx}(a). In contrast, $\widetilde{\Pi}_{\infty}$ requires either little modification or reduction, indicating the discrepancy in magnitude could arise from the model parameters for $\widetilde{\Pi} (\nu_*)$, but not from the scale size arguments.  

\section{Comparison over MAST database} \label{sec:compare}
\begin{figure}[!htbp]
\includegraphics[width=7.5 cm]{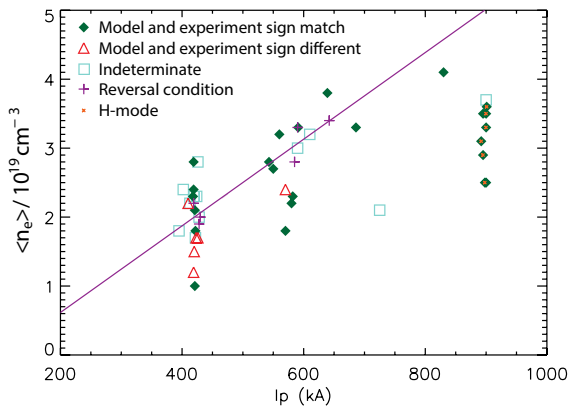}
\caption{\label{fig:database_panels} (Color online) Database of MAST intrinsic rotation comparing where the model and experiment agree and disagree for the sign of toroidal rotation.  Experimental conditions for reversals within shots are also shown, with the solid line being a linear fit to reversal conditions.  Data from H-mode plasmas are marked.}
\end{figure}

Figure~\ref{fig:database_panels} displays results from a database of Ohmic MAST shots, comparing the direction of the experimentally estimated toroidal rotation, typically at a radius $\sqrt{\psi}\approx 0.4-0.5$, to the predicted sign from the model at $\sqrt{\psi} = 0.5$.  The database includes both balanced double null, up-down symmetric and lower single null (LSN), up-down asymmetric discharges.  It also includes both L-mode and H-mode plasmas, with all the H-mode discharges at $I_p=900$ kA and marked in the figure.  Core electron temperatures ranged from about 350 eV at the lowest currents to about 900 eV at the highest, so temperature changes from improved confinement significantly impact the calculated collisionality, in concert with the density changes.  Conditions for reversals (when a reversal occurred within a discharge) are marked, with a linear fit shown.  Some cases, like t=280 ms in shot 29714 (see Figs.~\ref{fig:29714_traces}, ~\ref{fig:29714_coll}, and ~\ref{fig:29714_profile_mvx}) are taken to have an indeterminate sign, within experimental uncertainties of zero toroidal rotation (including estimates, available from~\cite{hillesheim_mastdbs_sub}, for systematic contributions to $v_{turb}$ other than $v_{\phi}$), but most times for comparison were chosen such that there was a definite sign outside estimates of total experimental uncertainties, which were conservatively bounded at $\sim 10$ km/s for $v_{\phi}$ for points of comparison.  In most cases in MAST, poloidal rotation is on the order of 1-2 km/s~\cite{field_comparison_2009} and consistent with neoclassical predictions, with discrepancies emerging near the magnetic axis in cases with internal transport barriers~\cite{field_comparison_2009,hillesheim_mastdbs_sub}, which are not relevant here.  Also from~\cite{hillesheim_mastdbs_sub}, even away from the ITB, differences between $v_{\phi}$ from charge exchange and $v_{turb} B/B_{\theta}$ from DBS were observed.  From the lower temperature gradient NBI-heated case in~\cite{hillesheim_mastdbs_sub} we chose the empirical uncertainty bound of $\sim 10$ km/s.  This is a safe estimate because the gradients in the Ohmic database are smaller than the gradients in the NBI-heated plasmas from ~\cite{hillesheim_mastdbs_sub} and as a result, the diamagnetic and phase velocity contributions should be smaller.  The model uncertainty is determined by calculating the boxcar ($\sim 20$ ms) average and standard deviation over an entire shot, as in Fig.~\ref{fig:29714_traces}(c).  An indeterminate sign is taken to be when the average is less than one standard deviation separated from zero.  The six cases where the model and experiment clearly disagreed (neither were indeterminate) are mostly clustered at low density and low current.

\begin{figure}[!htbp]
\includegraphics[width=8.5 cm]{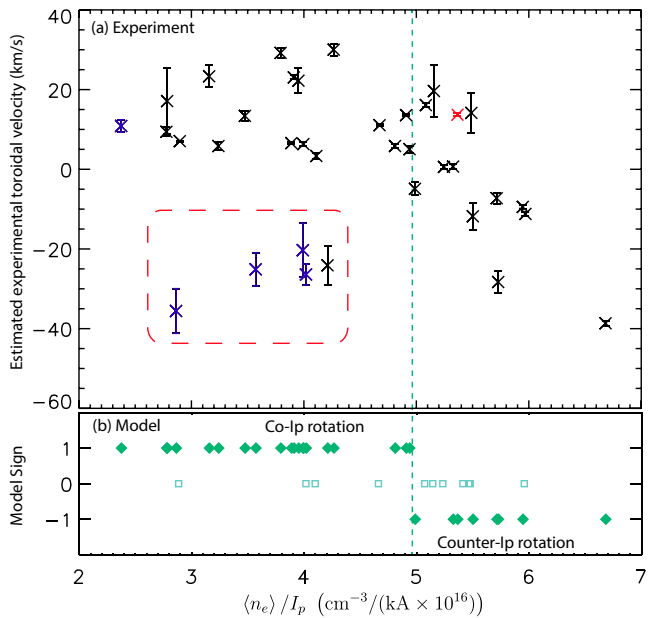}
\caption{\label{fig:global} (Color online) (a) Estimated core toroidal velocity, $v_{\phi} \approx v_{turb} B/B_{\theta}$, as a function of $\left< n_e \right>/I_p$. Cases with $I_p \approx 400$ kA and $\left< n_e \right> \leq 1.7 \times 10^{19}$ cm$^{-3}$ shown in blue.  The cases within the red dashed box and the red symbol are the cases that clearly disagreed, which are also plotted in red in Fig.~\ref{fig:database_panels}.  Positive is co-current rotation. (b) Sign of predicted toroidal rotation from the model for cases where the absolute value of the predicted velocity was larger than the uncertainty.  Cases where absolute value of predicted velocity was less than the uncertainty are plotted at zero as hollow boxes.  Vertical dashed line separates predicted co-rotation from predicted counter-rotation.}
\end{figure}

Figure~\ref{fig:global}(a) shows the experimental results over the full database with good DBS data, plotting the estimated core toroidal rotation against $\left< n_e \right>/I_p$, which here serves as an approximate global measure of collisionality.  The highest rotation cases at $v_{\phi} \approx 40$ km/s correspond to $M_{\phi} \approx 0.2$.  The data can be separated into two groups: Group \rom{1}, comprising the shots enclosed in the dashed box, and Group \rom{2}, which spans the rest of the database.  The cases in Group \rom{1} suggest that at low current and low density there is a second reversal branch, where a case like Fig.~\ref{fig:29714_traces} changes back to counter-rotation if the density drops further.  Most cases with $I_p \approx 400$ kA and $\left< n_e \right> \leq 1.7 \times 10^{19}$ cm$^{-3}$ belong to Group \rom{1}, and are plotted in blue in Fig.~\ref{fig:global}.  The remaining point in Group \rom{1} is a 600 kA shot, which might also be due to this second reversal, but there is insufficient data for a conclusion.  Notably, all five points in Group \rom{1} clearly disagree with the predictions of the model (the other case that clearly disagreed is plotted in red and occurs near the reversal condition for Group \rom{2}), while there is broadly good qualitative agreement with the rest of the data set.  There is a general trend for Group \rom{2}, composing most of the database, where the intrinsic rotation is co-$I_p$ at low  $\left< n_e \right>/I_p$ and becomes increasingly counter-$I_p$ at high $\left< n_e \right>/I_p$. This trend experimentally demonstrates the collisionality dependence.  Similar results relating intrinsic rotation reversals to global approximations for collisionality have also been observed in other experiments~\cite{rice_ohmic_2012,rice_nf_2013,reinke_density_2013}; however, this should only be taken as a very rough indicator since the intrinsic momentum flux should depend on local parameters.  

Figure~\ref{fig:global}(b) shows the predicted sign of core toroidal rotation, for cases where the absolute magnitude of the predicted velocity was larger than the uncertainty.  There is a clear transition marked by the vertical line, below which the model predicts co-$I_p$ rotation and above which the model predicts counter-$I_p$ rotation.  This shows the model reproduces a reversal density that scales with plasma current.  This demarcation is close to where the experimental points also start trending towards counter-$I_p$ rotation, providing strong evidence that the critical collisionality in the model is a good description of the experiment.  The difference in experimental behavior left and right of the demarcation is also qualitatively consistent with the model, where the limits at low and high collisionality are described separately.

We find good agreement over the database as a whole, with the model predicting the same sign as the experiment in about 80\% of cases where neither model nor experiment was indeterminate (about 30\% were indeterminate).  All cases where the model predicted the opposite rotation sign from the experiment were up-down symmetric L-mode plasmas; so, although larger discrepancies in magnitude were often found in up-down asymmetric LSN plasmas, this did not have a large impact on the comparisons of the sign of core rotation.  Discrepancies in sign near reversal conditions might simply be attributed to taking the model parameters from simulations of cyclone base case conditions, or to other approximations like $T_e=T_i$, $\Omega_{\phi}(1)=0$ rad/s, and $v_{\phi} \approx v_{turb} B/B_{\theta}$. There is a region at low density and low current where the model robustly predicts co-current toroidal rotation while counter-current rotation was observed in the experiment, due to a second reversal condition.  The second reversal branch cannot be explained by the simple model presented here, which could be due to the model neglecting additional effects that cause momentum transport or additional local parameters that can modify the the neoclassical distribution function.

\section{Discussion}
We have presented the first observations of spontaneous core intrinsic rotation transitions in a spherical tokamak, which demonstrates the apparent ubiquity of this phenomenon across a range of experiments.  This is unlike other transport properties such as the scaling of the energy confinement time, which is different in spherical and conventional tokamaks~\cite{kaye_prl_2007, valovic_2009}.  In contrast to other experiments, where the edge rotation is typically co-current and changes little, at MAST we observe that the edge rotation changes significantly in some cases, along with core.  

Two explanations for intrinsic rotation reversals have been proposed, both dependent on collisionality.  The first is that intrinsic rotation reversals are related to a transition in the linear turbulence drive, from ITG mode dominant to TEM dominant~\cite{angioni_intrinsic_2011,camenen_consequences_2011}.  Ostensibly, this is consistent with experiments investigating transitions in turbulence regimes, which have reported differences in turbulence characteristics correlated with collisionality~\cite{rettig_2001, garrard_2006, vermare_impact_2011, sung_cece_2013, arnichand_qcm_2014}; however, subsequent dedicated tests of this idea at Alcator C-mod~\cite{reinke_density_2013,white_multi-channel_2013}, KSTAR~\cite{shi_ech_2013}, and ASDEX-U~\cite{mcdermott_core_2014} have shown little concrete support.  The second is that neoclassical corrections, which depend on collisionality, to the equilibrium ion distribution function modify turbulent momentum transport~\cite{barnes_intrinsic_2013-1}.  We have described a simple 1D analytical model that captures the key physics of the latter for the purpose of testing the idea against experimental data. The comparisons revealed the model reproduces both the qualitative changes during rotation reversals in specific shots and the general trend of a reversal density that scales linearly with $I_p$.  Due to the robust nature of the model predictions for the sign of core toroidal rotation, the broad agreement over a range of experimental conditions along with the lack of free parameters fit to data are strong evidence that an important mechanism for explaining the rotation reversals in MAST has the characteristics that were used to construct the model: local intrinsic momentum flux that scales with the size of the diamagnetic effects $(B/B_{\theta})(\rho_i / L_{T_i})$, changes sign close to $\nu_* \approx 1$, and is independent of $\nu_*$ at high and low $\nu_*$. We do note that, particularly at low densities and currents in MAST, the comparison implies additional effects or parameters could also be important; similarly, the second reversal at high density reported in~\cite{mcdermott_core_2014} would require additional effects to explain.  

\acknowledgments
This work has been carried out within the framework of the EUROfusion Consortium and has received funding from the \textit{Euratom research and training programme} 2014-2018 under grant agreement No 633053, the RCUK Energy Programme under grant EP/I501045, and the US Department of Energy under DE-FG02-99ER54527.  To obtain further information on the data and models underlying this paper please contact PublicationsManager@ccfe.ac.uk.  The views and opinions expressed herein do not necessarily reflect those of the European Commission.  JCH's work partly supported through the Culham Fusion Research Fellowship and EFDA Fusion Research Fellowship programmes.  Dr. Terry Rhodes is gratefully acknowledged for allowing the Inspect spectral analysis program to be ported for use with MAST data.  Thanks to M. Carr for discussion of charge exchange recombination spectroscopy data.

\bibliographystyle{aipnum4-1}

\begin{thebibliography}{76}%
\makeatletter
\providecommand \@ifxundefined [1]{%
 \@ifx{#1\undefined}
}%
\providecommand \@ifnum [1]{%
 \ifnum #1\expandafter \@firstoftwo
 \else \expandafter \@secondoftwo
 \fi
}%
\providecommand \@ifx [1]{%
 \ifx #1\expandafter \@firstoftwo
 \else \expandafter \@secondoftwo
 \fi
}%
\providecommand \natexlab [1]{#1}%
\providecommand \enquote  [1]{``#1''}%
\providecommand \bibnamefont  [1]{#1}%
\providecommand \bibfnamefont [1]{#1}%
\providecommand \citenamefont [1]{#1}%
\providecommand \href@noop [0]{\@secondoftwo}%
\providecommand \href [0]{\begingroup \@sanitize@url \@href}%
\providecommand \@href[1]{\@@startlink{#1}\@@href}%
\providecommand \@@href[1]{\endgroup#1\@@endlink}%
\providecommand \@sanitize@url [0]{\catcode `\\12\catcode `\$12\catcode
  `\&12\catcode `\#12\catcode `\^12\catcode `\_12\catcode `\%12\relax}%
\providecommand \@@startlink[1]{}%
\providecommand \@@endlink[0]{}%
\providecommand \url  [0]{\begingroup\@sanitize@url \@url }%
\providecommand \@url [1]{\endgroup\@href {#1}{\urlprefix }}%
\providecommand \urlprefix  [0]{URL }%
\providecommand \Eprint [0]{\href }%
\@ifxundefined \urlstyle {%
  \providecommand \doi  [0]{\begingroup \@sanitize@url \@doi}%
  \providecommand \@doi [1]{\endgroup \@@startlink {\doibase
  #1}doi:\discretionary {}{}{}#1\@@endlink }%
}{%
  \providecommand \doi  [0]{doi:\discretionary{}{}{}\begingroup
  \urlstyle{rm}\Url }%
}%
\providecommand \doibase [0]{http://dx.doi.org/}%
\providecommand \Doi [0]{\begingroup \@sanitize@url \@Doi }%
\providecommand \@Doi  [1]{\endgroup\@@startlink{\doibase#1}\@@Doi}%
\providecommand \@@Doi [1]{#1\@@endlink}%
\providecommand \selectlanguage [0]{\@gobble}%
\providecommand \bibinfo  [0]{\@secondoftwo}%
\providecommand \bibfield  [0]{\@secondoftwo}%
\providecommand \translation [1]{[#1]}%
\providecommand \BibitemOpen [0]{}%
\providecommand \bibitemStop [0]{}%
\providecommand \bibitemNoStop [0]{.\EOS\space}%
\providecommand \EOS [0]{\spacefactor3000\relax}%
\providecommand \BibitemShut  [1]{\csname bibitem#1\endcsname}%











\bibitem{bortolon_observation_2006}
A. Bortolon \textit{et al}., Physical Review Letters 97, 235003 (2006).

\bibitem{rice_rotation_2011}
J. E. Rice \textit{et al}., Physical Review Letters 107, 265001 (2011).

\bibitem{angioni_intrinsic_2011}
C. Angioni \textit{et al}., Physical Review Letters 107, 215003 (2011).

\bibitem{garofalo_sustained_2002}
A. M. Garofalo \textit{et al}., Physical Review Letters 89, 235001 (2002).

\bibitem{burrell_effects_1997}
K. H. Burrell, Physics of Plasmas 4, 1499 (1997).




\bibitem{duval_bulk_2007}
B. P. Duval, \textit{et al}., Plasma Physics and Controlled Fusion 49, B195 (2007).

\bibitem{rice_observations_2011}
J. Rice \textit{et al}., Nuclear Fusion 51, 083005 (2011).


\bibitem{hillesheim_mastdbs_sub}
J.C. Hillesheim \textit{et al}., ``Measurement of high-k density fluctuation wavenumber spectrum in MAST and Doppler backscattering for spherical tokamaks,'' (submitted to Nucl. Fusion) \href{http://arxiv.org/abs/1407.2115}{arXiv:1407.2115 [physics.plasm-ph]}

\bibitem{lloyd_overview_2003}
 B. Lloyd \textit{et al}., Nuclear Fusion 43, 1665 (2003).



\bibitem{gurcan_2007}
\"{O}.D. Gurcan \textit{et al.}, Phys. Plasmas 14, 042306 (2007).

\bibitem{pat_2008}
P.H. Diamond \textit{et al.} Phys. Plasmas 15, 012303 (2008).

\bibitem{camenen_transport_2009}
Y. Camenen \textit{et al}., Physical Review Letters 102, 125001 (2009).




\bibitem{parra_turbulent_2010}
{F.I. Parra and P.J. Catto, Plasma Phys. Control. Fusion 52, 045004 (2010).}

\bibitem{parra_sources_2011}
{F.I. Parra \textit{et al}.,, Nucl. Fusion 51, 113001 (2011).}

\bibitem{waltz_gyrokinetic_2011}
{R.E. Waltz \textit{et al}.,, Phys. Plasmas 18, 042504 (2011).}

\bibitem{camenen_consequences_2011}
Y. Camenen, \textit{et al}., Nuclear Fusion 51, 073039 (2011).

\bibitem{sung_toroidal_2013}
{T. Sung \textit{et al}., Phys. Plasmas 20, 042506 (2013).}

\bibitem{lee_turbulent_2014}
{J.P. Lee \textit{et al}.,, Nucl. Fusion 54, 022002 (2014).}

\bibitem{lee_effect_2014}
{J.P. Lee \textit{et al}., Phys. Plasmas 21, 056106 (2014).}

\bibitem{barnes_intrinsic_2013-1}
M. Barnes \textit{et al}., Physical Review Letters 111, 055005 (2013).

\bibitem{felix_arxiv_2014}
F.I. Parra and M. Barnes, ``Intrinsic rotation in tokamaks. Theory.'' (\textit{accepted by Plasma Physics and Controlled Fusion}) \href{http://arxiv.org/abs/1407.1286}{arXiv:1407.1286 [physics.plasm-ph]}

\bibitem{rice_ohmic_2012}
J.E. Rice \textit{et al}., Physics of Plasmas 19, 056106 (2012)

\bibitem{reinke_density_2013}
M.L. Reinke \textit{et al}., Plasma Physics and Controlled Fusion 55, 012001 (2013).

\bibitem{mcdermott_core_2014}
R. M. McDermott \textit{et al}., Nuclear Fusion 54, 043009 (2014).



\bibitem{peeters_toroidal_2007}
{A.G. Peeters, C. Angioni and D. Strintzi, Phys. Rev. Lett. 98, 265003 (2007).}



\bibitem{casson_anomalous_2009}
{F.J. Casson \textit{et al}., Phys. Plasmas 16, 092303 (2009).}

\bibitem{barnes_turbulent_2011}
{M. Barnes \textit{et al}., Phys. Rev. Lett. 106, 175004 (2011).}

\bibitem{hinton_theory_1976}
F. L. Hinton and R. D. Hazeltine, Reviews of Modern Physics 48, 239 (1976).

\bibitem{helander_collisional_2005}
P. Helander and D. J. Sigmar, \textit{Collisional Transport in Magnetized Plasmas}, (Cambridge University Press, 2005).

\bibitem{dimits_comparisons_2000}
A. M. Dimits \textit{et al}., Physics of Plasmas 7, 969 (2000).








\bibitem{scannell_130_2010}
R. Scannell \textit{et al}., Review of Scientific Instruments 81, 10D520 (2010).

\bibitem{conway_high-throughput_2006}
N. J. Conway \textit{et al}., Review of Scientific Instruments 77, 10F131 (2006).





\bibitem{field_comparison_2009}
A.R. Field \textit{et al.}, Plasma Phys. Control. Fusion 51, 105002 (2009).

\bibitem{rice_nf_2013}
J.E. Rice \textit{et al}., Nucl. Fusion 53, 093015 (2013).

\bibitem{kaye_prl_2007}
S. M. Kaye \textit{et al.}, Phys. Rev. Lett. 98, 175002 (2007).

\bibitem{valovic_2009}
M. Valovi\v{c} \textit{et al.}, Nucl. Fusion 49, 075016 (2009).

\bibitem{rettig_2001}
C.L. Rettig \textit{et al.}, Phys. Plasmas 8, 2232 (2001). 

\bibitem{garrard_2006}
G.D. Conway \textit{et al.}, Nucl. Fusion 46, S799 (2006).  

\bibitem{vermare_impact_2011}
L. Vermare \textit{et al.}, Phys. Plasmas 18, 012306 (2011).

\bibitem{sung_cece_2013}
C. Sung \textit{et al.}, Nucl. Fusion 53, 083010 (2013).

\bibitem{arnichand_qcm_2014}
H. Arnichand \textit{et al.}, Nucl Fusion 54, 123017 (2014).

\bibitem{white_multi-channel_2013}
A. E. White \textit{et al}., Physics of Plasmas 20, 056106 (2013).

\bibitem{shi_ech_2013}
Y. J. Shi \textit{et al}., Nuclear Fusion 53, 113031 (2013).



\end{thebibliography}

\end{document}